\documentclass[pra,prl,twocolumn,superscriptaddress,floatfix,showpacs,10pt,letterpaper]{revtex4}
\usepackage{graphicx}
\usepackage{subfigure}
\usepackage{amssymb}
\usepackage{verbatim}
\usepackage{amsmath}
\usepackage{amscd}
\usepackage{amssymb}

\newcommand{\ket}[1]{|{#1}\rangle}
\newcommand{\bra}[1]{\langle{#1}|}

\newcommand{\ncd}{\newcommand}
\ncd{\QC}{$\mbox{QC}_{\cal{C}}\;$}
\ncd{\QCpr}{${\mbox{QC}_{\cal{C}}}^\prime\;$}
\ncd{\QCns}{$\mbox{QC}_{\cal{C}}$}
\ncd{\QCprns}{${\mbox{QC}_{\cal{C}}}^\prime$}
\ncd{\cskN}{{|\phi_{\{\kappa\} } \rangle}_{{\cal{C}}_N}}
\ncd{\cskNpr}{{|\phi_{\{\kappa^\prime\} } \rangle}_{{\cal{C}}_N}}
\ncd{\cskNtil}{{|\phi_{\{\tilde{\kappa} \} } \rangle}_{{\cal{C}}_N}}
\ncd{\csk}{{|\phi_{\{\kappa\} } \rangle}_{\cal{C}}}
\ncd{\csktil}{{|\phi_{\{\tilde{\kappa} \} } \rangle}_{\cal{C}}}
\ncd{\cskf}{|\phi_{\{\kappa\} } \rangle_{\cal{C}}}
\ncd{\csktilf}{|\phi_{\{\tilde{\kappa} \} } \rangle_{\cal{C}}}
\ncd{\bracsk}{\mbox{}_{\cal{C}}\langle\phi_{\{\kappa\} }|}
\ncd{\bracsktil}{\mbox{}_{\cal{C}}\langle\phi_{\{\tilde{\kappa} \} }|}
\ncd{\nbracsk}{\mbox{}_{\cal{C}}\langle\phi_{\{\kappa\} }}
\ncd{\nbracsktil}{\mbox{}_{\cal{C}}\langle\phi_{\{\tilde{\kappa} \} }}
\ncd{\cs}{|\phi \rangle_{\cal{C}}\;}
\ncd{\csns}{|\phi \rangle_{\cal{C}}}
\ncd{\nbgh}{\text{nbgh}}
\ncd{\Sab}{S^{ab}}
\ncd{\Sba}{S^{ba}}
\ncd{\ds}{\displaystyle}
\ncd{\ovl}{\overline}

\newtheorem{lemma}{Lemma}
\newtheorem{theorem}{Theorem}

\newcommand{\nc}{\newcommand}
\nc{\rnc}{\renewcommand}
\nc{\beq}{\begin{equation}}
\nc{\eeq}{{\end{equation}}}
\nc{\beqa}{\begin{eqnarray}}
\nc{\eeqa}{\end{eqnarray}}
\nc{\lbar}[1]{\overline{#1}}
\nc{\ketbra}[2]{|#1\rangle\!\langle#2|}
\nc{\braket}[2]{\langle#1|#2\rangle}
\nc{\proj}[1]{| #1\rangle\!\langle #1 |}
\nc{\avg}[1]{\langle#1\rangle}
\nc{\Rank}{\operatorname{Rank}}
\nc{\smfrac}[2]{\mbox{$\frac{#1}{#2}$}}
\nc{\Tr}{\operatorname{Tr}}
\nc{\id}{\operatorname{id}}
\nc{\1}{\openone}
\nc{\ox}{\otimes}
\nc{\dg}{\dagger}
\nc{\dn}{\downarrow}
\nc{\cA}{{\cal A}}
\nc{\cB}{{\cal B}}
\nc{\cC}{{\cal C}}
\nc{\cD}{{\cal D}}
\nc{\cE}{{\cal E}}
\nc{\cF}{{\cal F}}
\nc{\cG}{{\cal G}}
\nc{\cH}{{\cal H}}
\nc{\cI}{{\cal I}}
\nc{\cJ}{{\cal J}}
\nc{\cK}{{\cal K}}
\nc{\cL}{{\cal L}}
\nc{\cM}{{\cal M}}
\nc{\cN}{{\cal N}}
\nc{\cO}{{\cal O}}
\nc{\cP}{{\cal P}}
\nc{\cR}{{\cal R}}
\nc{\cS}{{\cal S}}
\nc{\cT}{{\cal T}}
\nc{\cX}{{\cal X}}
\nc{\cY}{{\cal Y}}
\nc{\cZ}{{\cal Z}}
\nc{\var}{\operatorname{var}}
\nc{\rar}{\rightarrow}
\nc{\lrar}{\longrightarrow}
\nc{\polylog}{\operatorname{polylog}}

\nc{\RR}{{{\mathbb R}}}
\nc{\CC}{{{\mathbb C}}}
\nc{\FF}{{{\mathbb F}}}
\nc{\NN}{{{\mathbb N}}}
\nc{\ZZ}{{{\mathbb Z}}}
\nc{\PP}{{{\mathbb P}}}
\nc{\QQ}{{{\mathbb Q}}}
\nc{\UU}{{{\mathbb U}}}
\nc{\EE}{{{\mathbb E}}}
\nc{\Icoh}{{I^{\rm coh}}}
\nc{\Qca}{{Q_{\rm ss}}}
\nc{\Qcaa}{{Q^{(1)}_{\rm ss}}}
\nc{\Dcaa}{{D^{(1)}_{{\rm ss}\rightarrow}}}
\nc{\Dca}{{D_{{\rm ss}\rightarrow}}}

\nc{\be}{\begin{equation}}
\nc{\ee}{{\end{equation}}}
\nc{\bea}{\begin{eqnarray}}
\nc{\eea}{\end{eqnarray}}
\nc{\<}{\langle}
\rnc{\>}{\rangle}
\nc{\Hom}[2]{\mbox{Hom}(\CC^{#1},\CC^{#2})}
\nc{\rU}{\mbox{U}}

\begin{document}

\title{Quantum Capacity Approaching Codes for the Detected-Jump Channel}

\author{Markus Grassl}
\affiliation{Centre for Quantum
  Technologies, National University of Singapore, Singapore 117543, Singapore} 

\author{Zhengfeng Ji}
\affiliation{Perimeter Institute for Theoretical Physics, Waterloo, ON, N2L2Y5, Canada}
\affiliation{State Key Laboratory of Computer Science, Institute of
  Software, Chinese Academy of Sciences, Beijing, China}
  
\author{Zhaohui Wei}
\affiliation{Centre for Quantum
  Technologies, National University of Singapore, Singapore 117543, Singapore}

\author{Bei Zeng} \affiliation{Department of Mathematics $\&$
  Statistics, University of Guelph, Guelph, Ontario, Canada}
\affiliation{Institute for Quantum Computing and Department of
  Combinatorics $\&$ Optimization, University of Waterloo, Waterloo,
  Ontario, Canada} \date{\today}

\begin{abstract}
The quantum channel capacity gives the ultimate limit for the rate at
which quantum data can be reliably transmitted through a noisy quantum
channel.  Degradable quantum channels are among the few channels
whose quantum capacities are known.  Given the quantum capacity of a
degradable channel, it remains challenging to find a practical coding
scheme which approaches capacity. 
Here we discuss code designs for the detected-jump channel, a
degradable channel with practical relevance describing the physics of
spontaneous decay of atoms with detected photon emission.  
We show that this channel can be used to simulate a binary classical
channel with both erasures and bit-flips. The capacity of the
simulated classical channel gives a lower bound on the quantum
capacity of the detected-jump channel.  When the jump probability is
small, it almost equals the quantum capacity.  Hence using a classical
capacity approaching code for the simulated classical channel yields a
quantum code which approaches the quantum capacity of the
detected-jump channel.
\end{abstract}

\maketitle %\narrowtext

Information theory was founded by Claude E. Shannon in 1948.  In his
landmark paper \cite{shannon-1948} he defined the notion of channel
capacity, which is a tight upper bound on the amount of information
that can be reliably transmitted over a noisy communication
channel. The capacity of a channel is given by a single numerical
quantity, characterizing the amount of information that can be
transmitted asymptotically per channel use.  Shannon also showed that
it is possible to encode messages in such a way that the number of
extra bits transmitted is as small as possible.  Unfortunately his
proof does not give any explicit recipe for these optimal
codes.  After decades of efforts the goal of finding explicit codes
which reach the limits predicted by Shannon's original work has been
achieved~\cite{mocoding}.

The origins of quantum information theory can be seen in the early
1990s.  One of the questions addressed is the problem of reliable
transmission of information through a quantum
channel~\cite{nielsenchuang}. Unlike the situation for classical
channels, one can define several capacities for quantum channels.  The
capacity of a quantum channel depends on the auxiliary resources
allowed, the class of protocols allowed, and whether the information
to be transmitted is classical or quantum~\cite{shor-2003,Smith-2010}.
In this paper we discuss in particular the quantum capacity of a
quantum channel which gives the ultimate bound on the rate at which
quantum data can be reliably transmitted through a noisy quantum
channel.  This capacity is also called the one-way quantum capacity,
where all communication is directly from the sender to the receiver
over the noisy quantum channel.

Given a quantum channel $\Phi$, the quantum capacity of the channel is
given by~\cite{beth-2008,shor-2003,Smith-2010}
\begin{equation}
Q_C(\Phi)=\lim_{n\rightarrow\infty}\frac{1}{n}I^{\text{coh}}(\Phi^{\otimes n}),
\end{equation}
where $I^{\text{coh}}$ is the coherent information of a channel
defined by
\begin{equation}
I^{\text{coh}}(\Phi)=\max_{\rho}\left(S(\Phi(\rho))-S(\Phi^{C}(\rho))\right).
\end{equation}
Here $S(\rho)$ is the von Neumann entropy of $\rho$, and $\Phi^{C}$ is
the complementary channel of $\Phi$.

Degradable quantum channels are among the few classes of channels
whose quantum capacities are known. A channel $\Phi$ is degradable if
there is another channel $\Psi$ such that
\begin{equation}
\Psi\circ\Phi=\Phi^{C}.
\end{equation}
It was
shown in~\cite{dev-2003} that the capacity of a degradable channel
satisfies 
\begin{equation}
Q_C(\Phi)=I^{\text{coh}}(\Phi),
\label{eq:capacity}
\end{equation}
so that the capacity can
be computed~\cite{wolf-small}.  Given the quantum capacity of a
degradable channel, the quantum coding theorem guarantees the
existence of a scheme to encode the message such that the capacity can
be achieved. However, it remains a challenge to find a practical
coding scheme which approaches the capacity. 

%MG PRL does not like claims of priority
%MG To our knowledge this important issue has not yet been addressed in
%MG literature.

In this letter we discuss code designs for the detected-jump channel,
which is a degradable channel with practical relevance describing the
physics of spontaneous decay of atoms with detected photon emission.
Our main observation is that in the Hadamard basis, the detected-jump
channel simulates a binary classical channel with both erasure and
flip. This allows us to use classical codes to serve as the basis for
a quantum code, resulting in good quantum jump codes with parameters
better than those of any previously known codes.  The capacity of the
simulated classical channel also gives a lower bound for the quantum
capacity of the detected-jump channel. When the jump probability is
small, the scheme reaches almost the quantum capacity.  Hence using
classical codes approaching capacity for the simulated classical
channel as the basis for the quantum codes results in quantum codes
that approach the quantum capacity of the detected-jump channel.

\textit{Detected-Jump Channel} The detected-jump channel was
considered in~\cite{alber-01}. It is a channel with practical
relevance describing the physics of spontaneous decay of atoms with
detected photon emission.  The spontaneous decay is traditionally
described by the jump channel (also called the amplitude damping
channel, denoted by $\Phi_{AD}$) described by the Kraus operators
\begin{equation}
A_0=\begin{pmatrix} 1 & 0 \\0 &
\sqrt{1-\gamma} \end{pmatrix}
\quad\text{and}\quad A_1=\begin{pmatrix} 0 &
\sqrt{\gamma} \\0 & 0 \end{pmatrix},
\label{eq:ADKraus}
\end{equation}
that is,
\begin{equation}
\Phi_{AD}(\rho)=\sum_iA_i\rho A_i^{\dagger},
\end{equation}
If the photon emission of the spontaneous decay can be detected, then
the corresponding channel, called the detected-jump channel (denoted
by $\Phi_{DJ}$) is given by
\begin{equation}
\Phi_{DJ}(\rho)=\sum_i\left(A_i\rho A_i^{\dagger}\right)_{\text{sys}}\otimes \ket{i}\bra{i}_{\text{aux}}.
\label{eq:DJKraus}
\end{equation}
The complementary channel of $\Phi_{DJ}$ is given by 
\begin{equation}\label{eq:DJ_comp}
\Phi_{DJ}^{C}(\rho)=\sum_i\text{Tr}(A_i\rho A_i^{\dagger})\ket{i}\bra{i}_{\text{aux}}.
\end{equation}
It is easy to see from (\ref{eq:DJKraus}) and (\ref{eq:DJ_comp}) that
$\Phi_{DJ}^{C}$ can be obtained from $\Phi_{DJ}$ by taking the trace
of the first system, so $\Phi_{DJ}$ is degradable. Therefore, the quantum
capacity of $\Phi_{DJ}$ can be directly calculated using Eq. (\ref{eq:capacity}).

\begin{figure}[htbp]
  \centerline{%
  \begin{picture}(180,100)(-20,-10)
  \put(5,0){\makebox(0,0)[r]{$\ket{+}\equiv 0$}}
  \put(5,80){\makebox(0,0)[r]{$\ket{-}\equiv 1$}}
  \put(95,0){\makebox(0,0)[l]{$0\equiv  \ket{+}_{\text{sys}}\ket{0}_{\text{aux}}$}}
  \put(95,40){\makebox(0,0)[l]{$E\equiv \ket{1}_{\text{sys}}\ket{1}_{\text{aux}}$}}
  \put(95,80){\makebox(0,0)[l]{$1\equiv \ket{-}_{\text{sys}}\ket{0}_{\text{aux}}$}}
  \put(50,-2){\makebox(0,0)[t]{\footnotesize$1-p_E-p_F$}}
  \put(10,0){\vector(1,0){80}}
  \put(10,0){\vector(2,1){80}}
  \put(50,16){\makebox(0,0)[t]{\footnotesize$p_E$}}
  \put(10,0){\vector(1,1){80}}
  \put(30,30){\makebox(0,0)[t]{\footnotesize$p_F$}}
  \put(10,80){\vector(1,0){80}}
  \put(10,80){\vector(2,-1){80}}
  \put(50,64){\makebox(0,0)[b]{\footnotesize$p_E$}}
  \put(10,80){\vector(1,-1){80}}
  \put(30,50){\makebox(0,0)[b]{\footnotesize$p_F$}}
  \put(50,82){\makebox(0,0)[b]{\footnotesize$1-p_E-p_F$}}
  \end{picture}}
  \caption{A binary classical channel $\Xi$ derived from the
    quantum channel $\Phi'_{DJ}$ by measuring the system in the basis
    $\{\ket{+},\ket{-}\}$ and the auxiliary system in the basis $\{\ket{0},\ket{1}\}$.}
  \label{fig:cc}
\end{figure}
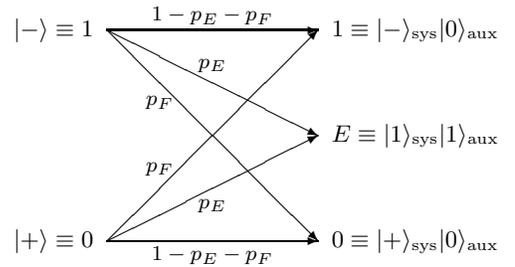

\textit{Error-Correcting Codes for the Detected-Jump Channel} The
construction of quantum error-correcting codes for $\Phi_{DJ}$ has
been discussed in~\cite{alber-01,alber-03,ahn-03,ahn-04}.  Here we
provide a new observation for the code construction. Starting from
Eq.~(\ref{eq:DJKraus}), applying a controlled-NOT operation from the
auxiliary system to the system, we obtain an equivalent channel
$\Phi'_{DJ}$ given by
\begin{equation}
\Phi'_{DJ}(\rho)=\sum_i\left(A'_i\rho {A'_i}^{\dagger}\right)_{\text{sys}}\otimes \ket{i}\bra{i}_{\text{aux}},
\end{equation}
where
\begin{equation}
A'_0=A_0=\begin{pmatrix} 1 & 0 \\0 &
\sqrt{1-\gamma} \end{pmatrix}
\quad\text{and}\quad A'_1=\begin{pmatrix} 0 &
0 \\0 & \sqrt{\gamma} \end{pmatrix}.
\label{eq:ADKraus2}
\end{equation}
As both $A'_0$ and $A'_1$ are diagonal, the channel with these
operators actually simulates some classical channel in the Hadamard
basis.  We clarify this fact by the following lemma.
\begin{lemma}
In the Hadamard basis, $\Phi'_{DJ}$ simulates a binary classical
channel $\Xi$ with both erasure and bit flip, given by
Fig.~\ref{fig:cc}.  The erasure probability is $p_E=\frac{\gamma}{2}$
and the bit flip probability is given by
$p_F=\left[\frac{1}{2}(1-\sqrt{1-\gamma})\right]^2$.
\end{lemma}
\textbf{Proof}: Note that $A'_1\ket{+}=\sqrt{\frac{\gamma}{2}}\ket{1}$
and $A'_1\ket{-}=-\sqrt{\frac{\gamma}{2}}\ket{1}$, i.e., in the
Hadamard basis $A'_1$ erases all information, but at the same time
this is indicated by the auxiliary state $\ket{1}_{\text{aux}}$.  The
probability for this event is $p_E=\frac{\gamma}{2}$.  At the same
time, in this basis $A_0'$ can be expressed as
\begin{equation}
A'_0=\left[\frac{1}{2}(1+\sqrt{1-\gamma})\right]I+\left[\frac{1}{2}(1-\sqrt{1-\gamma})\right]Z
\end{equation} 
Hence, when no error was indicated by the auxiliary qubit and if we
measure the system qubit in the Hadamard basis, the qubit will be
flipped with probability $p_F=[\frac{1}{2}(1-\sqrt{1-\gamma})]^2$, or
will not change with probability $1-p_E-p_F$. $\square$

This lemma allows us to use classical codes for the classical channel
$\Xi$ in the Hadamard basis to serve as the basis for the
quantum codes for the equivalent quantum channels $\Phi_{DJ}$ and
$\Phi'_{DJ}$.

% Yet it is unclear what is the good classical code $\Xi$ with
% general $p_E,p_F$. We show in the special case of
% $p_F=\left[\frac{1}{2}(1-\sqrt{1-\gamma})\right]^2$, we have a simple
% construction.

In the following we construct a quantum error-correcting code
$\mathcal{Q}$ which is a subspace of $(\CC^2)^{\otimes n}$, the space of
$n$ qubits.  Recall that for a $K$-dimensional code space spanned by
the orthonormal basis states $\ket{\psi_i}$, $i=1,\ldots,K$ and a set
of errors $\cE$ there is a physical operation correcting all elements
$E_\mu \in \cE$ if the error correction conditions
\cite{bennett-1996-54,KL97} are satisfied:
\begin{equation}
\forall_{i,j,\mu,\nu}\quad\bra{\psi_i} E_\mu^\dag E_\nu \ket{\psi_j} =
\alpha_{\mu\nu}\delta_{ij},
\label{eq:qecccondition}
\end{equation}
where $\alpha_{\mu \nu}$ depends only on $\mu$ and $\nu$. 
% If the matrix $(\alpha_{\mu \nu})$ has full rank the code is said to
% be nondegenerate, otherwise it is degenerate.  Instead of the
conditions (\ref{eq:qecccondition}) one can consider approximate
error-correction \cite{leung-1997}, i.e. Eq.~(\ref{eq:qecccondition})
is only fulfilled up to a certain order $t$.

\begin{theorem}
For an $(n,K,d)$ classical code $C$, the residual error probability
over the channel $\Xi$ is of order $O(\gamma^d)$.  From the
classical code $C$, one obtains an $((n,K))$ quantum code which
corrects errors of the quantum channel $\Phi_{DJ}$ up to order
$O(\gamma^t)$ where $t=d-1$.
\end{theorem}
\textbf{Proof}: An $(n, K, d)$ classical code $C$ can simultaneously
correct $e$ erasures and $f$ errors provided $e+2f<d$ (see, e.g.,
\cite[Theorem 1.11.6]{HuPl03}). Note that
$p_E=\frac{\gamma}{2}$ and
\begin{equation}
p_F=\left[\frac{1}{2}(1-\sqrt{1-\gamma})\right]^2=\frac{1}{16}\gamma^2+O(\gamma^3).
\end{equation}
Therefore, using the code $C$, the residual error probability over the
channel $\Xi$ is improved to $O(\gamma^d)$.

Similarly, in order to correct errors of $\Phi_{DJ}$ up to order
$O(\gamma^t)$, one needs to find an appropriate quantum code such that
the error-correction conditions (\ref{eq:qecccondition}) hold up to
order $O(\gamma^t)$ \cite{leung-1997,Cedric,Prabha}.

% Suppose $\mathbf{x},\mathbf{y}\in C$.  We know that the Hamming
% distance between them is not smaller than $d$. 

Based on the classical code $C$, we construct a corresponding quantum
code $\mathcal{Q}$ with basis
\begin{equation}
\{\ket{\psi_{\mathbf{x}}}=H^{\otimes n}\ket{\mathbf{x}}\colon\mathbf{x}\in C\}.
\label{eq:defQ}
\end{equation}
Now we prove that this code is sufficient for our task.

Setting $A=X+iY$ and $B=I-Z$, where $X,Y,Z$ are Pauli operators for qubits, we have
\begin{equation}
A_1=\frac{\sqrt{\gamma}}{2}A\quad\text{and}\quad A_0=I-\frac{\gamma}{4}B+O(\gamma^2).
\end{equation}
It can be shown that in order to improve the fidelity of the
transmission through a detected-jump channel from $1-\gamma$ to
$1-\gamma^{t+1}$, it is sufficient to satisfy the error-correction
conditions Eq.~(\ref{eq:qecccondition}) for $e$ $A$-errors and $f$
$B$-errors with $e+2f\le t$ \cite[Section 8.7]{thesis:gottesman}.  By
the construction given in Eq.~(\ref{eq:defQ}), this is equivalent to
requiring that the classical code $\Xi$ corrects $e$ erasure errors
and $f$ bit flip errors (note here $d=t+1$). $\square$

This theorem yields good quantum jump codes with parameters better
than those previously known in~\cite{alber-01,alber-03,ahn-03,ahn-04}.

\textit{Capacity Approaching Codes for the Detected-Jump Channel}
Contrary to the classical situation~\cite{mocoding}, there is almost
nothing known about the construction of practical, quantum capacity
approaching codes for quantum channels, not even for degradable
channels.  However, as a naive example consider a quantum channel
$\Theta$ with Kraus operators $\sqrt{1-p}I$ and $\sqrt{p}X$.  It is
straightforward to calculate that the quantum capacity of this channel
is given by $H(p)$, where $H$ is the binary entropy.  The
corresponding classical channel is the binary symmetric channel with
bit flip probability $p$, whose capacity is well-known to be $H(p)$,
too.  This is not a surprise.  Although we use $\Theta$ to transmit
quantum information, its behavior is exactly classical.  Apparently,
one can use capacity approaching codes for the binary symmetric
channel as a basis for a quantum code for $\Theta$, which then
approaches the quantum capacity of $\Theta$.

We borrow the idea from this naive example to design quantum capacity
approaching codes for the detected-jump channel.  Due to the previous
discussion, we can use the classical codes for the simulated channel
$\Xi$ as a basis for a quantum code for the quantum channel
$\Phi_{DJ}$. The capacity of the classical channel $\Xi$ gives a lower
bound for the quantum capacity of the quantum channel
$Q_C(\Phi_{DJ})$.  

As $\Phi_{DJ}$ is degradable, its capacity can be
computed using Eq.~(\ref{eq:capacity}).  It is given by
\begin{alignat}{6}
Q_C(\Phi_{DJ})=\max_{x\in[0,1]}\{&(1-\gamma x)\log(1-\gamma x)\nonumber\\[-1.5ex]
&-(1-x)\log(1-x)\nonumber\\
&-(1-\gamma)x\log(1-\gamma)x\}.
\end{alignat}

Recall that for the transmission of classical information over a
classical channel, the capacity is given by the maximal mutual
information
\begin{equation}
\sup_{p_X}I(X;Y)
\end{equation}
between the input $X$ and the output $Y$, where the maximization is
over all input distributions $p_X$ \cite{info}.  The classical
capacity of the simulated classical channel $\Xi$ turns out to be
\begin{eqnarray}
H(p_E,\frac{1-p_E}{2},\frac{1-p_E}{2})-H(p_E,p_F,1-p_E-p_F).
\end{eqnarray}

\begin{figure}[htbp]
  \includegraphics[width=2.5in]{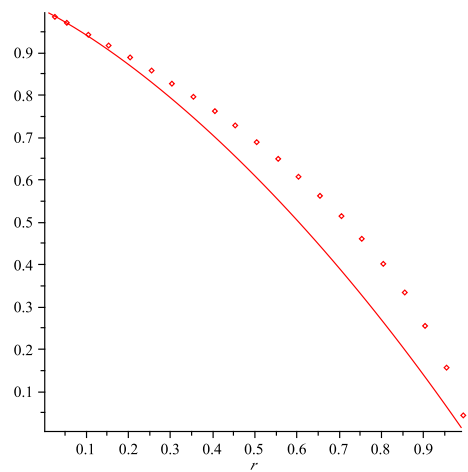}
  \caption{The capacities for the region $\gamma\in[0,1]$. The line is
    the capacity of the classical channel $\Xi$, and the dots
    give the capacity of the quantum channel $\Phi_{DJ}$.}
  \label{fig:cap1}
\end{figure}

The capacities of the classical channel $\Xi$ and the quantum channel
$\Phi_{DJ}$ are plotted in Fig.~\ref{fig:cap1}.

\begin{figure}[htbp]
  \includegraphics[width=2.5in]{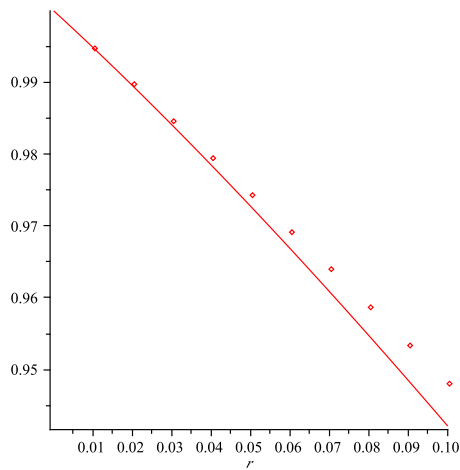}
  \caption{The capacities for the region $\gamma\in[0,0.1]$. The line
    is the capacity of the classical channel $\Xi$, and the dots give
    the capacity of the quantum channel $\Phi_{DJ}$.}
  \label{fig:cap2}
\end{figure}

From Fig.~\ref{fig:cap1} one can see that the capacity of the
simulated classical channel gives a good lower bound on the quantum
capacity of the detected-jump channel.  When the jump probability is
small, the lower bound almost equals the quantum capacity, see
Fig.~\ref{fig:cap2}.  Hence using a classical capacity approaching
code for the simulated classical channel yields a quantum code which
approaches the quantum capacity of the detected-jump channel.

\textit{Summary and Discussion} We use classical codewords as basis
for a quantum code to construct good quantum codes for the
detected-jump channel with parameters better than those of any
previously known codes. Our method gives a lower bound on the quantum
capacity of the detected-jump channel.  When the jump probability is
small, it almost equals the quantum capacity.  Hence using a classical
capacity approaching code for the simulated classical channel yields a
quantum code which approaches the quantum capacity of the
detected-jump channel.

Our discussion is closely related to that of environment assisted
channels~\cite{werner-04,smolin-05}. However, our detected-jump
channel given in Eq.~(\ref{eq:DJKraus}) is different from an environment
assisted channel of the amplitude damping channel given in
Eq.~(\ref{eq:ADKraus}). Because we only allow the access of the
environment in a fixed basis, the quantum capacity of $\Phi_{DJ}$ is
lower than the environment assisted capacity of the amplitude damping
channel.

Our result provides the first example of capacity-approaching codes of
degradable quantum channels. We hope that our method sheds light on
the problem of finding capacity-approaching codes for other degradable
quantum channels.

\textbf{Acknowledgement} We thank C{\'e}dric B{\'e}ny, Runyao Duan,
and Andreas Winter for helpful discussions. ZJ and BZ thank the Centre
for Quantum Technologies for hospitality. Centre for Quantum
Technologies is a Research Centre of Excellence funded by the Ministry
of Education and the National Research Foundation of Singapore. ZJ
acknowledges support from NSF of China (Grant Nos.~60736011 and
60721061); his research at Perimeter Institute is supported by the
Government of Canada through Industry Canada and by the Province of
Ontario through the Ministry of Research $\&$ Innovation. 
ZW acknowledges grant from the Centre for Quantum Technologies, 
and the WBS grant under contract no. R-710-000-008-271. BZ is
supported by NSERC and CIFAR.

%%%%%%%%%%%%%%%%%%%%%%%%%%%%%%%%%%%%%%%%%%%%%%%%%%%%%%%%%%%%%%%%%%%%%%%%%%%%%
\bibliography{CAAD}

\end{document}